\documentclass[%
 reprint,
 amsmath,amssymb,
 aps,
]{revtex4-1}

\usepackage{graphicx}
\usepackage{dcolumn}
\usepackage{bm}

\usepackage[top=0.8in, bottom=0.8in, left=0.7in, right=0.7in]{geometry}
\usepackage{fancyhdr}

\usepackage{lastpage}

\begin{document}

\preprint{APS/123-QED}

\title{V-type electromagnetically induced transparency and saturation effect at the gas-solid interface}

\author{Tengfei Meng}
\affiliation{State Key Laboratory of Quantum Optics and Quantum Optics Devices, Institute
of Laser spectroscopy, Shanxi University, Taiyuan 030006, People's Republic
of China}

\author{Yanting Zhao}
\thanks{Corresponding author: zhaoyt@sxu.edu.cn}
\affiliation{State Key Laboratory of Quantum Optics and Quantum Optics Devices, Institute
of Laser spectroscopy, Shanxi University, Taiyuan 030006, People's Republic
of China}

\author{Zhonghua Ji}
\affiliation{State Key Laboratory of Quantum Optics and Quantum Optics Devices, Institute
of Laser spectroscopy, Shanxi University, Taiyuan 030006, People's Republic
of China}

\author{Dianqiang Su}
\affiliation{State Key Laboratory of Quantum Optics and Quantum Optics Devices, Institute
of Laser spectroscopy, Shanxi University, Taiyuan 030006, People's Republic
of China}

\author{Liantuan Xiao}
\affiliation{State Key Laboratory of Quantum Optics and Quantum Optics Devices, Institute
of Laser spectroscopy, Shanxi University, Taiyuan 030006, People's Republic
of China}

\author{Suotang Jia}
\affiliation{State Key Laboratory of Quantum Optics and Quantum Optics Devices, Institute
of Laser spectroscopy, Shanxi University, Taiyuan 030006, People's Republic
of China}

\date{\today}

\begin{abstract}
We theoretically study electromagnetically induced transparency (EIT) in reflection spectra of V-type system at the gas-solid interface. In addition to a narrow dip arising from the EIT effect, we find the other particular saturation effect induced by pump field, which does not exist in $\Lambda$\ or $\Xi$\ -type system reflection spectra. The saturation effect only induces an intensity decrement in the reflection spectra, and there is no influence on the narrow dip arising from the EIT effect. We detailedly calculate and analyze the dependence of V-type system reflection spectra on probe field intensity, pump field intensity, coherent decay rate, and the initial population after the collision between atoms and the interface.
\end{abstract}

\maketitle

\section{\label{sec:level1}INTRODUCTION}

Over the past two decades, many researchers are interested in atomic electromagnetically induced transparency (EIT) phenomena \cite{Harris} relevant to the nonlinear efficiency\cite{Deng2001,Fleischhauer2005}. EIT has been applied to many fields such as slow light \cite{Kasapi,Hau,Hang}, magneto-optical switch \cite{McGloin}, quantum information storage \cite{Fleisehhauer2000,Liu,Lukin}, and huge Kerr effect \cite{Deng2007}.

Gas-solid interface has been widely studied as a physical resource \cite{Bordo2003,Bordo2004,Arora,Tao,Itami}. Recently, many physics phenomena such as EIT\cite{Thomas,Du}, Autler-Townes splitting \cite{Sautenkov} and transient four-wave mixing \cite{Lorenz} are demonstrated experimentally in reflection spectra at the gas-solid interface. Due to the abundant physics at the interface, it is meaningful to make an investigation on EIT at the interface through the reflection spectra arising from a thin layer of atoms near the boundary. The collision between atoms and the interface can give rise to a sub-Doppler structure for reflection spectra because of the symmetry breaking in the velocity distribution of optically polarized atoms near the interface \cite{Zhao,Laliotis}. Thus, reflection spectroscopy is known as an excellent tool to study the properties of atoms in the vicinity of the interface \cite{Stern,Segundo}.

EIT has been studied in $\Lambda,$\ $\Xi$\ and V-type systems in free space \cite{Li1995,Banacloche,Lazoudis}, and has also been studied in $\Lambda$\  and $\Xi$\ systems at the interface \cite{Nienhuis1994,Kampen}. However, there is no report about EIT in V-type system at the interface as far as we know. V-type system has an essential difference compared to $\Lambda$\ or $\Xi$\ -type systems, which has been studied in reference \cite{Lazoudis}. For a three-level $\Lambda$\ system, atoms are mainly populated in two ground states. When $\Lambda$\ -type system couples with a weak probe field and a strong pump field, almost all atoms are populated in the ground state coupling with probe field. Hence, there is no saturation in $\Lambda$\ -type system. For a three-level $\Xi$\ -type system, if pump field couples with middle state and excited state, there is also no saturation. Because most atoms are populated in the ground state coupling with probe field. Saturation shows up in V-type EIT system, and it is because that pump field modifies the atomic population of ground state and exited state. Thus, the V-type reflection spectrum is influenced by the EIT and saturation when the system couples with a very weak probe field and a strong pump field. This is very different from $\Lambda$\ and $\Xi$\ -type systems.

In this paper, we study the V-type system EIT by reflection spectra at the interface. The paper is organized as follows. In Sec.\ref{AA}, we discuss the theory of two-level atomic reflection spectra. In Sec.\ref{BB}, we perform the calculations based on Liouville equations, and analyze V-type system EIT of reflection spectra. In Sec.\ref{CC}, we analyze EIT and saturation based on the narrow dip near the resonance of reflection spectra. In Sec.\ref{DD}, we study the dependence of V-type system reflection spectra on probe field intensity, pump field intensity, coherent decay rate and the initial population after the collision between atoms and the interface. Finally, we summarize the main results in Sec.\ref{EE}.

\section{\label{AA}GENERAL THEORY FOR REFLECTION SPECTRUM}

\begin{figure}[h]
\begin{tabular}{cc}
\includegraphics[width=3.4in]{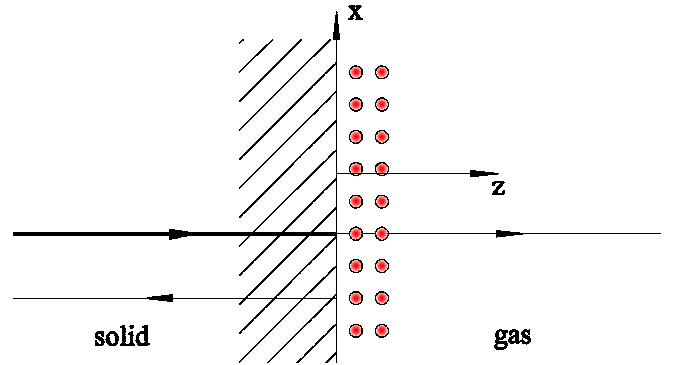}
\end{tabular}
\caption{(Color online) The geometry of gas-solid interface with optical field at normal incidence.
}
	\label{fig21}
\end{figure}

 Fig. \ref{fig21} is the theoretical model which is similar as in Ref. \cite{Nienhuis1988}. We consider the interface between the atomic vapor and a dielectric material with a real refractive index n. The interface is taken as the x-y plane. The atomic vapor fills the half space with $z>0$\, and the dielectric fills the half space with $z<0.$\ When a plane wave is incident on the interface one part of the field is reflected back into the solid dielectric, and the other part of the field refract into the vapor. The refracted field in the vapor drives the active atoms, thereby creating a dipole polarization. This polarization emits radiation in the direction of the reflected wave, thereby modifying the reflection coefficient. In the case of normal incidence, the reflection coefficient can be written as
\begin{equation}
{R}=
 \left(\frac{n-1}{n+1}\right)^2
  + \frac{4n(n-1)}{(n+1)^3}ReT.
 \label{eq:a}
\end{equation}
The first item of Eq.(\ref{eq:a}) is the intensity reflection coefficient at the interface in absence of the atomic vapor. The second item is the intensity reflection coefficient caused by the polarized atoms. ReT is the real part of a dimensionless quantity T, and the parameter T can be expressed as
\begin{equation}
{T}=
 \frac{ik}{\varepsilon_0\textbf{E}_1}\int_0^\infty{e^{2ikz}}{{\textbf{P}_0}(z)}{dz},
 \label{eq:b}
\end{equation}
where $\textbf{E}_1$\ is the amplitude of the radiation field propagating through the interface, $k$\ is the wave number of the radiation field, ${\textbf{P}_0}{(z)}$\ is the intensity of the polarization caused by the radiation field through the interface, and it is determined by the expected value of atomic dipole moment which is dependent on atomic position $z$\ and velocity $v.$\ It can be expressed as
\begin{equation}
{\textbf{P}_0(z)}=
 \int_{-\infty}^\infty{NW(v)}{\textbf{p}_0(z,v)}{dv}.
 \label{eq:c}
\end{equation}
Here $N$\ is the atomic number density, $W(v)$\ is a normalized Maxwell-Boltzmann distribution function and ${\textbf{p}_0}{(z,v)}$\ is the individual atomic dipole moment. An atom with a positive value of $v$\ has left the surface a time $z/v$\ ago, the collision between the atom and interface induces the atomic de-excitation and changes the states of the atom. On the other hand, when $v$\ is negative, the atom has traveled many lifetimes in the field, so that it will reached its steady state, and the atomic dipole moment is independent on the position. Therefore, we can get
\begin{equation}
{{\textbf{p}_0}(z,v)}=
 \overline{\textbf{p}}_0(v).
 \label{eq:d}
\end{equation}
The dipole moment obeys the relation
 \begin{equation}
{{\textbf{p}_0}(z,v)}=
 {\textbf{u}_{12}}{\sigma_{21}{(v,z)}}.
 \label{eq:e}
\end{equation}
Here ${\textbf{u}}_{12}$\ is the transition dipole moments and $\sigma_{21}{(z,v)}$\  is the non-diagonal element of the reduced density matrix $\sigma.$\

According to the Eqs.(\ref{eq:b})-(\ref{eq:e}) and the Laplace transform of $\sigma_{21}$\ \cite{Nienhuis1988}, we get
 \begin{widetext}
 \begin{equation}
{T}=
 -\frac{N\textbf{u}_{21}}{2{\varepsilon_0}\textbf{E}_1}({\int_{-\infty}^0}{W(v){\overline{\sigma}}_{21}(v)}{dv}+{\int_0^{+\infty}}{W(v)(-2ik){\widehat{\sigma}}_{21}(-2ik,v)}{dv}).
 \label{eq:g}
\end{equation}
\end{widetext}
In this expression $\overline{\sigma}_{21}(v)$\ is the stationary value of $\sigma_{21},$\ and $\widehat{\sigma}_{21}(-2ik,v)$\ is the Laplace transform of $\sigma_{21}.$\ The reflection coefficient $R_{SR},$\ which represents the contribution due to the presence of the polarized atoms arising from Eqs.(\ref{eq:a}) and (\ref{eq:g}), is expressed as
\begin{equation}
{R_{SR}}=
 \frac{4n(n-1)}{(n+1)^3}{\frac{N\textbf{u}_{21}}{2{\varepsilon_0}\textbf{E}_1}}\phi.
 \label{eq:h}
\end{equation}
The parameter $\phi={\phi_+}+{\phi_-}$\ is the sum of the real dimensionless parameters $\phi_-$\ and $\phi_+,$\ which is defined by
 \begin{subequations}
\begin{eqnarray}
{\phi_+}&=&
 -Re{\int_0^{+\infty}}{W(v)(-2ik){\widehat{\sigma}}_{21}(-2ik,v)}{dv}, \label{apa}
\\
{\phi_-}&=&
 -Re{\int_{-\infty}^0}{W(v){\overline{\sigma}}_{21}(v)}{dv}. \label{apb}
\end{eqnarray}
\end{subequations}
Here $\phi_+$\ and $\phi_-$\ represent the contribution to the reflection spectrum from atoms with positive and negative velocities respectively.

\section{\label{BB}THEORETICAL MODEL OF V-TYPE THREE-LEVEL SYSTEM}

\begin{figure}[h]
\includegraphics[width=3.4in]{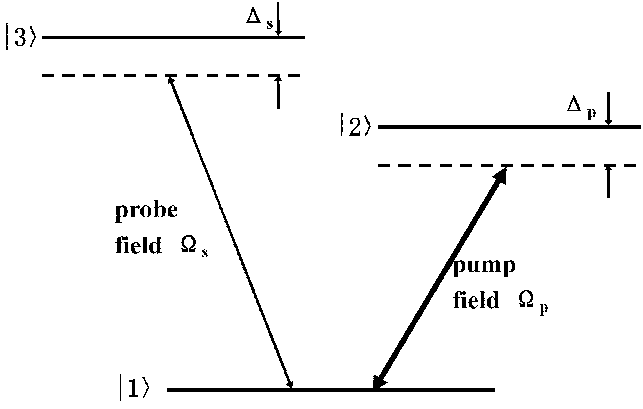}
\caption{V-type three-level system. The pump field couples the lower level $|1\rangle$\ and the upper level $|2\rangle.$\ The probe field couples the lower level $|1\rangle$\ and the upper level $|3\rangle.$\ $\Omega_s$\ and $\Omega_p$\ are probe and pump Rabi frequency respectively. $\Delta_s$\ and $\Delta_p.$\ are the frequency detuning of probe and pump fields.
}
	\label{fig1}
\end{figure}

V-type three-level system is shown in Fig.\ref{fig1}, and it contains a lower level $|1\rangle$\  and two upper levels $|2\rangle$\  and $|3\rangle.$\ The probe field couples level $|1\rangle$\ and $|3\rangle.$\ The pump field couples level $|1\rangle$\ and $|2\rangle.$\ EIT is induced by the coherence of level $|2\rangle$\ and $|3\rangle.$\ $\Omega_p$\ and  $\Omega_s$\ are the Rabi frequencies of probe field and pump field. We consider that the pump and probe fields propagate in opposite directions.

The atomic state of V-type system can be described by the density matrix $\rho({v_z},z,t),$\ which obeys the Liouville equation \cite{Schuller}. The equations can be written in the time-independent form if we introduce a reduced density matrix $\sigma$\ transformed from density matrix $\rho({v_z},z,t),$\
\begin{subequations}
\begin{eqnarray}
{\rho_{21}}&=&
 e^{-i({\omega_p}t-{k_p}z)}\sigma_{21}, \label{appda}
\\
{\rho_{31}}&=&
 e^{-i({\omega_s}t-{k_s}z)}\sigma_{31}, \label{appdb}
\\
{\rho_{32}}&=&
 e^{-i({(\omega_s-\omega_p)}t-{(k_s-k_p)}z)}\sigma_{32}. \label{appdc}
\end{eqnarray}
\end{subequations}
Here $\omega_p$\ and $k_p$\ are the frequency and wave number of pump field. $\omega_s$\ and $k_s$\ are the frequency and wave number of probe field. Then we obtain equations below in the rotating-wave approximation
\begin{subequations}
\begin{eqnarray}
v_z{\frac{d}{dz}}\sigma_{11}&=&\frac{i\Omega_p}{2}{(\sigma_{21}-\sigma_{12})}
+\frac{i\Omega_s}{2}{(\sigma_{31}-\sigma_{13})}\nonumber \\
 & &+\Gamma_2\sigma_{22}+\Gamma_3\sigma_{33}, \label{appaa}
\\
{v_z{\frac{d}{dz}}\sigma_{22}}&=&
 -\frac{i\Omega_p}{2}{(\sigma_{21}-\sigma_{12})}-\Gamma_2\sigma_{22}, \label{appab}
\\
{v_z{\frac{d}{dz}}\sigma_{33}}&=&
 -\frac{i\Omega_s}{2}{(\sigma_{31}-\sigma_{13})}-\Gamma_3\sigma_{33}, \label{appac}
\\
{v_z{\frac{d}{dz}}\sigma_{21}}&=&
 \frac{i\Omega_p}{2}{(\sigma_{11}-\sigma_{22})}
 {-\frac{i\Omega_s}{2}\sigma_{23}}+(i\widetilde{\Delta}_p-\gamma_{21})\sigma_{21}, \nonumber \\
 & &\label{appad}
 \\
{v_z{\frac{d}{dz}}\sigma_{31}}&=&
 \frac{i\Omega_s}{2}{(\sigma_{11}-\sigma_{33})}
 {-\frac{i\Omega_p}{2}\sigma_{32}}
+(i\widetilde{\Delta}_s-\gamma_{31})\sigma_{31}, \nonumber \\
 & &\label{appae}
 \\
 {v_z{\frac{d}{dz}}\sigma_{32}}&=&
 \frac{i\Omega_s}{2}{\sigma_{12}}
 {-\frac{i\Omega_p}{2}\sigma_{31}}
 +[(i(\widetilde{\Delta}_s-\widetilde{\Delta}_p)-\gamma_{32}]\sigma_{32},\nonumber \\
 & &\label{appaf}
\end{eqnarray}
\end{subequations}
where $\widetilde{\Delta}_s$\ and $\widetilde{\Delta}_p$\ are the Doppler-shift frequency detunings of the pump and probe fields respectively,

\begin{subequations}
\begin{eqnarray}
{\widetilde{\Delta}_s}=
 {\omega_s}-{\omega_{31}}-{k_s}v=\Delta_s-{k_s}v, \label{appea}
\\
{\widetilde{\Delta}_p}=
 {\omega_p}-{\omega_{21}}-{k_p}v=\Delta_p-{k_p}v. \label{appeb}
\end{eqnarray}
\end{subequations}
Here $\omega_{ij}(i{\neq}j, i,j=1,2,3)$\ is the transition frequency from level $|i\rangle$\ to level $|j\rangle.$\ $\Gamma_i$\ is the population decay rate of level $|i\rangle.$\

 \begin{equation}
{\gamma_{ij}}=
 \frac{{\Gamma_i}+{\Gamma_j}}{2}+\gamma_{ijp}.
 \label{eq:k}
\end{equation}
Here $\gamma_{ijp}{(i{\neq}j)}$\ denotes the coherence decay rate between levels $|i\rangle$\ and $|j\rangle$\ induced by the collision between
atoms and the interface \cite{Fulton}. In this formula,
\begin{subequations}
\begin{eqnarray}
\gamma_{12p}&=&q, \label{appa}
\\
\gamma_{13p}&=&q, \label{appb}
\\
\gamma_{23p}&=&2q, \label{appc}
\end{eqnarray}
\end{subequations}
where $q$\ represents the relative decay rate.

In order to solve Eq.(\ref{appaa})-(\ref{appaf}), we introduce the Laplace transformation
 \begin{equation}
{\widehat{\sigma}(v_z,p)}=
\int_0^{+\infty}dz{e^{-pz}\sigma(v_z,z)}.
 \label{eq:l}
\end{equation}
The derivatives on the left-hand side of Eq.(\ref{appaa})-(\ref{appaf})are then easily obtained from the relation

 \begin{equation}
\int_0^{+\infty}{e^{-pz}\frac{d}{dz}\sigma}=
-\sigma(z=0)+p{\widehat{\sigma}}.
 \label{eq:m}
\end{equation}
Then $\sigma(v_z,z)$\ is determined by the set of algebraic equations

\begin{subequations}
\begin{eqnarray}
{v_z}p\widehat{\sigma}_{11}&=&\frac{i\Omega_p}{2}{(\widehat{\sigma}_{21}-\widehat{\sigma}_{12})}
+\frac{i\Omega_s}{2}{(\widehat{\sigma}_{31}-\widehat{\sigma}_{13})}\nonumber \\
 & &+\Gamma_2\widehat{\sigma}_{22}+\Gamma_3\widehat{\sigma}_{33}+\sigma_{11}(z=0), \label{appba}
\\
{v_z}p\widehat{\sigma}_{22}&=&
 -\frac{i\Omega_p}{2}{(\widehat{\sigma}_{21}-\widehat{\sigma}_{12})}
 -\Gamma_2\widehat{\sigma}_{22}+\sigma_{22}(z=0), \label{appbb}
\\
{v_z}p\widehat{\sigma}_{33}&=&
 -\frac{i\Omega_s}{2}{(\widehat{\sigma}_{31}-\widehat{\sigma}_{13})}
 -\Gamma_3\widehat{\sigma}_{33}+\sigma_{33}(z=0), \label{appbc}
\\
{v_z}p\widehat{\sigma}_{21}&=&
 \frac{i\Omega_p}{2}{(\widehat{\sigma}_{11}-\widehat{\sigma}_{22})}
 {-\frac{i\Omega_s}{2}\widehat{\sigma}_{23}}+(i\widetilde{\Delta}_p-\gamma_{21})\widehat{\sigma}_{21}, \nonumber \\
 & &\label{appbd}
 \\
{v_z}p\widehat{\sigma}_{31}&=&
 \frac{i\Omega_s}{2}{(\widehat{\sigma}_{11}-\widehat{\sigma}_{33})}
 {-\frac{i\Omega_p}{2}\widehat{\sigma}_{32}}
+(i\widetilde{\Delta}_s-\gamma_{31})\widehat{\sigma}_{31}, \nonumber \\
 & &\label{appbe}
 \\
{v_z}p\widehat{\sigma}_{32}&=&
 \frac{i\Omega_s}{2}{\widehat{\sigma}_{12}}
 {-\frac{i\Omega_p}{2}\widehat{\sigma}_{31}}
 +[i(\widetilde{\Delta}_s-\widetilde{\Delta}_p)-\gamma_{32}]\widehat{\sigma}_{32},\nonumber \\
 & &\label{appbf}
\end{eqnarray}
\end{subequations}
where $p$\ is equal to $-2ik_s.$\ These equations are used to get $\widehat{\sigma}_{31}$ which is related to the V-type EIT reflection spectrum ${\phi_+}$\ for atoms with positive velocity. The initial state is normalized as
 \begin{equation}
\sigma_{11}{(z=0)}+\sigma_{22}{(z=0)}+\sigma_{33}{(z=0)}=
1.
 \label{eq:n}
\end{equation}
For atoms with negative velocity the stationary value of $\sigma$\ is obtained from the relation
 \begin{equation}
\overline{\sigma}=
\lim_{p\rightarrow0}{p\widehat{\sigma}({v_z,p})}.
 \label{eq:o}
\end{equation}
The corresponding reduced density matrix equations are
\begin{subequations}
\begin{eqnarray}
&&\frac{i\Omega_p}{2}{(\overline{\sigma}_{21}-\overline{\sigma}_{12})}
+\frac{i\Omega_s}{2}{(\overline{\sigma}_{31}-\overline{\sigma}_{13})}+\Gamma_2\overline{\sigma}_{22}+\Gamma_3\overline{\sigma}_{33}=0,\nonumber\\& &\label{appca}
\\
&&-\frac{i\Omega_p}{2}{(\overline{\sigma}_{21}-\overline{\sigma}_{12})}-\Gamma_2\overline{\sigma}_{22}=0, \label{appcb}
\\
&&-\frac{i\Omega_s}{2}{(\overline{\sigma}_{31}-\overline{\sigma}_{13})}-\Gamma_3\overline{\sigma}_{33}=0, \label{appcc}
\\
&&\frac{i\Omega_p}{2}{(\overline{\sigma}_{11}-\overline{\sigma}_{22})}
 {-\frac{i\Omega_s}{2}\overline{\sigma}_{23}}+(i\widetilde{\Delta}_p-\gamma_{21})\overline{\sigma}_{21}=0, \label{appcd}
 \\
&&\frac{i\Omega_s}{2}{(\overline{\sigma}_{11}-\overline{\sigma}_{33})}
 {-\frac{i\Omega_p}{2}\overline{\sigma}_{32}}
+(i\widetilde{\Delta}_s-\gamma_{31})\overline{\sigma}_{31}=0,\label{appce}
 \\
&&\frac{i\Omega_s}{2}{\overline{\sigma}_{12}}
 {-\frac{i\Omega_p}{2}\overline{\sigma}_{31}}
 +[(i(\widetilde{\Delta}_s-\widetilde{\Delta}_p)-\gamma_{32}]\overline{\sigma}_{32}=0.\label{appcf}
\end{eqnarray}
\end{subequations}
The equations above are used to get the $\overline{\sigma}_{31}$\ which is related to the V-type EIT reflection spectrum $\phi_-$\ for atoms with negative velocity. Thus, V-type EIT reflection spectrum can be obtained by the superposition of $\phi_+$\ and $\phi_-.$\

\section{\label{CC}ELECTROMAGNETICALLY INDUCED TRANSPARENCY AND SATURATION EFFECT}
As described in Ref. \cite{Lazoudis}, EIT and saturation occur simultaneously in a V-type system in free space, which is induced by pump field. In order to analyze EIT and saturation in reflection spectra at the interface, we study the reduced density matrix equations (\ref{appba})-(\ref{appbf})  and (\ref{appca})-(\ref{appcf}). If $\sigma_{32}$\ is set as zero, the coherence effects involving the interplay of the two laser fields can be turned off effectively(i.e., with $\sigma_{32}=0,$\ then $\sigma_{21}$\ and $\sigma_{31}$\ can be written simply in terms of the population differences $\sigma_{11}-\sigma_{22}$\ and $\sigma_{11}-\sigma_{33},$\ respectively). Therefore, by setting $\sigma_{32}=0$ and then solving the density matrix equations of motion, we can study the effects of saturation by itself and distinguish them from those caused by EIT.

From Eq. (\ref{appbd}) and under the condition of $\sigma_{32}=0$,\ we obtain the $\widehat{\sigma}_{21}$\ coherence after the collision between atoms and the interface.
\begin{equation}
\widehat{\sigma}_{21}=
\frac{1}{p{v_z}-{i\widetilde{\Delta}_p+\gamma_{21}}}\frac{i\Omega_p}{2}(\widehat{\sigma}_{11}-\widehat{\sigma}_{22}).
 \label{eq:p}
\end{equation}
Similarly, from Eq. (\ref{appbe}) the $\widehat{\sigma}_{31}$\ coherence is given as
\begin{equation}
\widehat{\sigma}_{31}=
\frac{1}{p{v_z}-{i\widetilde{\Delta}_s+\gamma_{31}}}\frac{i\Omega_s}{2}(\widehat{\sigma}_{11}-\widehat{\sigma}_{33}).
 \label{eq:q}
\end{equation}
From Eqs. (\ref{appcd}) and (\ref{appce}) we can also get $\overline{\sigma}_{21}$\ and  $\overline{\sigma}_{31}$\ before the collision between atoms and the interface.
\begin{equation}
\overline{\sigma}_{21}=
\frac{1}{-{i\widetilde{\Delta}_p+\gamma_{21}}}\frac{i\Omega_p}{2}(\overline{\sigma}_{11}-\overline{\sigma}_{22}).
 \label{eq:r}
\end{equation}

 \begin{equation}
\overline{\sigma}_{31}=
\frac{1}{-{i\widetilde{\Delta}_s+\gamma_{31}}}\frac{i\Omega_s}{2}(\overline{\sigma}_{11}-\overline{\sigma}_{33}).
 \label{eq:s}
\end{equation}
We use (\ref{appba})-(\ref{appbf}), (\ref{appca})-(\ref{appcf}) and (\ref{eq:p})-(\ref{eq:s}) to analyze reflection spectra.

\begin{figure}[h]
\includegraphics[width=3.4in]{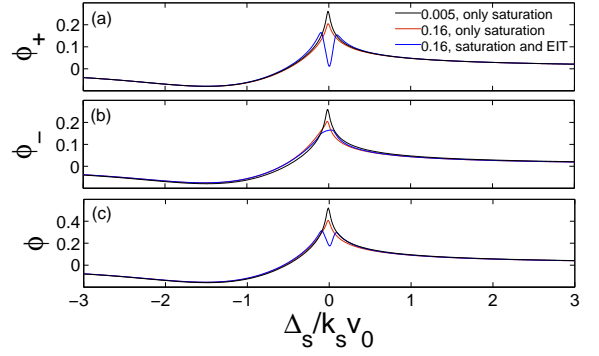}
\label{fig3}
\caption{(Color online) The V-type reflection spectra with and without EIT. (a) reflection spectra for atoms with positive velocity. (b) reflection spectra for atoms with negative velocity. (c) reflection spectra for atoms with positive and negative velocity. The black (red) lines show the case only considering the saturation effect when pump Rabi frequency is $0.005{k_s}v_0$\ $(0.16{k_s}v_0).$\ The blue lines show the case considering saturation and EIT when pump Rabi frequency is $0.16{k_s}v_0.$\ The other parameters are $\Omega_s=0.005{k_s}v_0, \Delta_p=0, \sigma_{11}=1, \sigma_{22}=0, \sigma_{33}=0, \Gamma_1=0, \Gamma_2=0.03{k_s}v_0,\Gamma_3=0.03{k_s}v_0,{k_p}v_0=-{k_s}v_0$\ and $q=0.$\ ${k_s}v_0$\ denotes the Doppler width of reflection spectrum and $v_0$\ is the most probable thermal velocity. }
	\label{fig3}
\end{figure}

Fig. \ref{fig3}. shows V-type system reflection spectra for atoms with positive and negative velocity. The black lines show the case only considering the saturation effect when  pump Rabi frequency is $0.005{k_s}v_0.$\ The red lines show the case only considering the saturation effect when pump Rabi frequency is $0.16{k_s}v_0.$\ Both considering saturation and EIT, the blue lines show the case when pump Rabi frequency is $0.16{k_s}v_0.$\ By compared the black line to the red line in Fig.3(a) and (b) we can find that the peak decreases when pump Rabi frequency increases for both positive and negative atoms. The increment of pump Rabi frequency enhances saturation, which reduces the population of $|1\rangle.$\ Then the coherence between level $|1\rangle$\  and level $|2\rangle$\ decreases. The decrement of the coherence between level $|1\rangle$\  and level $|2\rangle$\ decreases the reflectivity, then the peak near the resonance decreases.

By compared the red line to the blue line as shown in Fig. \ref{fig3}(a) we can find that the narrow dip near the resonance is induced by EIT, which is caused by the coherence between level $|2\rangle$\ and level$|3\rangle.$\ Fig. \ref{fig3}(b) shows that the peak is reduced when $\sigma_{32}\neq0$,\ and it is caused by the coherence between level $|2\rangle$\ and level$|3\rangle.$\  Fig. \ref{fig3}(c) is the superposition of reflection spectra for atoms with positive and negative velocity. A dip appears near the resonance is caused by EIT which is not related to saturation. By comparison, saturation and EIT cause the dip respectively in free space\cite{Lazoudis}. So the contributions of saturation and EIT in V-type reflection spectra are different from those spectra in free space.

\section{\label{DD}THE PARAMETERS DEPENDENCE OF V-TYPE REFLECTION SPECTRA}
In this section, we calculate and analyze the dependence of V-type system reflection spectra on probe Rabi frequency, pump Rabi frequency, coherence decay rate and the initial population after the collision between atoms and the interface. The parameters in the simulation are set as follows: $\Gamma_1=0,$\ $\Gamma_2=0.03{k_s}v_0,$\ $\Gamma_3=0.03{k_s}v_0$\ and ${k_p}v_0=-{k_s}v_0.$\ Reflection spectra with and without the collision between atoms and the interface are considered separately.

\subsection{V-type EIT spectra dependence on probe Rabi frequency}
\begin{figure}[h]
\begin{tabular}{cc}
\includegraphics[width=3.4in]{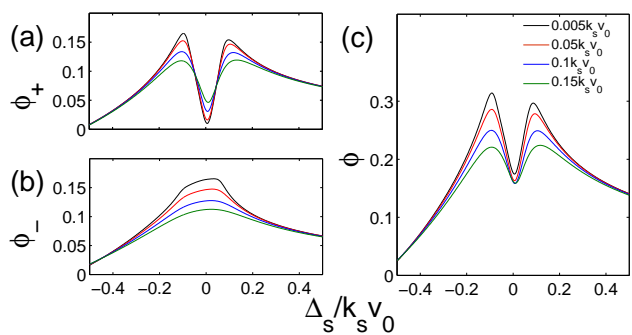}
\end{tabular}
\caption{(Color online) The V-type EIT reflection spectra under different probe fields. (a) reflection spectra for atoms with positive velocity. (b) reflection spectra for atoms with negative velocity. (c) the superposition of reflection spectra for atoms with positive and negative velocity. The black, red, blue, green lines show the cases when probe Rabi frequency is $0.005k_s{v_0},$\ $0.05k_s{v_0},$\ $0.1k_s{v_0}$\ and $0.15k_s{v_0}$\ respectively. The pump Rabi frequency is $0.16k_s{v_0},$\ and the other parameters are the same as that in Fig. \ref{fig3}.
 }
	\label{fig4}
\end{figure}
Fig. \ref{fig4} shows the V-type EIT reflection spectra under different probe Rabi frequencies $\Omega_s.$\ Fig. \ref{fig4}(a) displays reflection spectra for atoms with positive velocity. The increment of $\Omega_s$\ enhances saturation between level $|1\rangle$\ and level $|3\rangle$, which leads to the weaker coherence between level $|1\rangle$\ and level $|3\rangle.$\  The two peaks on both sides of dips decrease, and it is caused by saturation between level $|1\rangle$\ and level $|3\rangle$\ . The decrement of population in level $|1\rangle$\ and the increment of population in level $|3\rangle$\ also reduce EIT caused by the coherence between level $|2\rangle$\ and level $|3\rangle.$\ Therefore the dip caused by EIT decreases.  Fig. \ref{fig4}(b) shows reflection spectra for atoms with negative velocity. Saturation and three-level coherence reduce the reflectivity significantly near the resonance as $\Omega_s$\ increases. Fig. \ref{fig4}(c) shows the superposition of reflection spectra for atoms with positive and negative velocity. The main change are the two peaks on both sides of dip, and there is little change caused by EIT.

With the increment of $\Omega_s,$\ the population of level $|1\rangle$\ decreases and the population of level $|3\rangle$\ increases until complete saturation. During this process, reflection spectra decrease near the resonance if we neglect pump field \cite{Nienhuis1988}. When $\Omega_s$\ increases, the number of atoms which participate in saturation increases in atomic ensemble. Thus the number of atoms which participate in the coherence decreases. The spectral profile decreases near the resonance as in Fig. \ref{fig4}(c). So reflection spectra in three-level system change as same as the two-level system when we increase probe Rabi frequency.

\subsection{V-type EIT spectra dependence on pump Rabi frequency}

\begin{figure}[h]
\begin{tabular}{cc}
\includegraphics[width=3.4in]{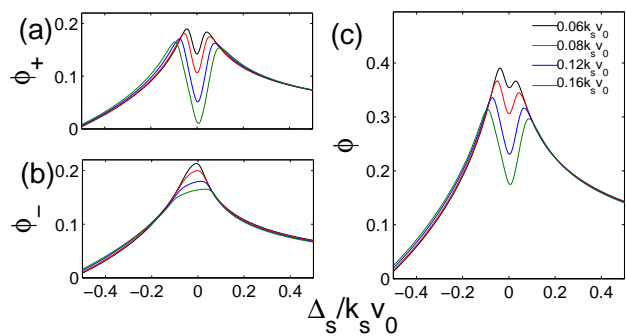}
\end{tabular}
\caption{(Color online) The V-type EIT reflection spectra under different pump fields. (a) reflection spectra for atoms with positive velocity. (b) reflection spectra for atoms with negative velocity. (c) the superposition of reflection spectra for atoms with positive and negative velocity. The black, red, blue, green lines show the cases when pump Rabi frequency is $0.06k_s{v_0},$\ $0.08k_s{v_0},$\ $0.12k_s{v_0}$\ and $0.16k_s{v_0}$ respectively. The other parameters are the same as that in Fig. \ref{fig3}. }
	\label{fig5}
\end{figure}

Fig. \ref{fig5} displays the dependence of the V-type EIT reflection spectra on the pump Rabi frequency $\Omega_p.$\ Fig. \ref{fig5}(a) shows the V-type EIT spectra for atoms with positive velocity. The increment of $\Omega_p$\ enhances the coherence between level $|2\rangle$\ and level $|3\rangle$. Hence, the dip caused by EIT increases with the enhancement of pump field. With the increment of $\Omega_p,$\ the population of level $|1\rangle$\ decreases, and the coherence between level $|1\rangle$\ and level $|3\rangle$ reduces. It induces the decrement of the reflectivity near the resonance. Thus, the two peaks on both sides of dip decrease simultaneously. It is caused by the enhanced saturation between level $|1\rangle$\ and level $|3\rangle.$ Fig. \ref{fig5}(b) shows the V-type EIT reflection spectra for atoms with negative velocity. Saturation between level $|1\rangle$\ and level $|3\rangle$ and the coherence between level $|2\rangle$\ and level $|3\rangle$ reduce the reflectivity significantly near the resonance as $\Omega_p$\ increases. Fig. \ref{fig5}(c) shows the superposition of reflection spectra for atoms with positive and negative velocity. The main changes happen at the dip and the two peaks both sides of dip. The dip is changed by EIT, and the two peaks both sides of dips are changed by saturation.

\subsection{V-type EIT spectra dependence on coherence decay rate}

\begin{figure}[h]
\begin{tabular}{cc}
\includegraphics[width=3.4in]{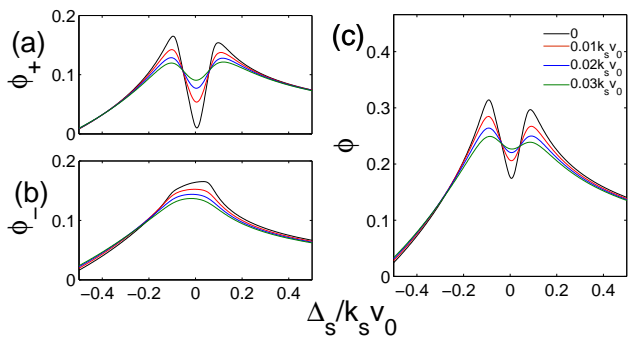}
\end{tabular}
\caption{(Color online) The V-type EIT reflection spectra under different coherence decay rates $\gamma_{ijp}$\ relying on relative decay rate $q$\ (a) reflection spectra for atoms with positive velocity. (b) reflection spectra for atoms with negative velocity. (c) the superposition of reflection spectra for atoms with positive and negative velocity. The black, red, blue, green lines show the cases when $q$\ is $0,$\ $0.01k_s{v_0},$\ $0.02k_s{v_0}$\ and  $0.03k_s{v_0}$ respectively. The pump Rabi frequency is $0.16k_s{v_0},$\ and the other parameters are the same as that in Fig. \ref{fig3}.
 }
	\label{fig6}
\end{figure}

The collision between atoms should be considered due to the dense atomic vapor near the interface \cite{Sautenkov,Wang}. Fig. \ref{fig6} displays the V-type EIT reflection spectra on condition of different coherence decay rates $\gamma_{ijp},$\ which is induced by the collision between atoms.

Fig. \ref{fig6}(a) shows the V-type EIT reflection spectra for atoms with positive velocity. The depth of the dip decreases and the two peaks on both sides of dip decrease  when $\gamma_{ijp}$\ increases. The increment of $\gamma_{ijp}$\ causes the decrement of the coherence between level $|1\rangle$\ and level $|3\rangle$, then the two peaks  decrease. The increment of $\gamma_{ijp}$\ causes the decrement of the coherence between level $|2\rangle$\ and level $|3\rangle$, then the depth of the dip decreases. Fig. \ref{fig6}(b) shows the V-type EIT reflection spectra for atoms with negative velocity. The reflectivity decreases near the resonance with the increment of the coherence decay rate, which is caused by the decrement of the coherence between level $|1\rangle$\ and level $|3\rangle$ and decrement of between level $|2\rangle$\ and level $|3\rangle$. Fig. \ref{fig6}(c) shows the superposition of reflection spectra for atoms with positive and negative velocity. The two peaks and the depth of the dip decrease with the increment of $\gamma_{ijp},$\ and they are corresponding to decrement of the coherence between level $|1\rangle$\ and level $|3\rangle$ and the coherence between level $|2\rangle$\ and level $|3\rangle$. The depth of dip decrease when $\gamma_{ijp}$\ increases, which is consistent with the spectra in free space.

\subsection{V-type EIT spectra dependence on initial population}

\begin{figure}[h]
\begin{tabular}{cc}
\includegraphics[width=3.4in]{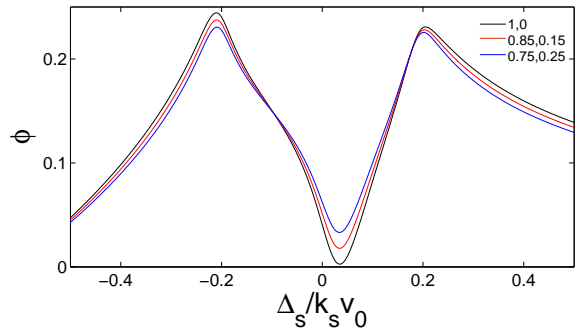}
\end{tabular}
\caption{(Color online) The V-type EIT reflection spectra which is the superposition of reflection spectra for atoms with positive and negative velocity. The V-type EIT reflection spectra have different initial population of level $|2\rangle$\ for atoms with positive velocity. The black (red,blue)line represents that the populations of level $|1\rangle$\ and level $|2\rangle$\ are $\sigma_{11}(z=0)=1(0.85,0.75), \sigma_{22}(z=0)=0(0.15,0.25).$\  The Rabi frequency of pump field is ${\Omega_p}=0.4{k_s}v_0,$\ and the other parameters are the same as that in fig. \ref{fig3}.
 }
	\label{fig7}
\end{figure}

Atoms are in a steady state before the collision between atoms and the interface, and the population can be calculated by (\ref{appca})-(\ref{appcf}). When we calculate the population by the parameters as same as in Fig. \ref{fig3}, the consequence is that the population of level $|1\rangle$\ is approximate 0.75 and level $|2\rangle$\ is approximate 0.25. There is no population in level $|3\rangle.$\ The collision between atoms and the interface causes the de-excitation of level $|2\rangle.$\ The change of population in level $|2\rangle$\ is different when atoms collide with different dielectric material. The state of atoms after the collision between atoms and the interface is assumed the initial state for atoms with positive velocity. The dependence of initial population is shown in Fig. \ref{fig7}. It can be seen that the depth of the dip increases with the decrement of $\sigma_{11}(z=0).$\  Then the decrement of $\sigma_{11}(z=0)$\ decreases EIT in different z, and it causes the decrement of the depth of the dip.

\section{\label{EE}CONCLUSIONS}

In this paper, we theoretically study the V-type EIT reflection spectra by density matrix equations. Through the numerical results of V-type EIT spectra at the interface, we find a narrow dip which is related to EIT. We also find a saturation effect which does not exist in reflection spectra of $\Lambda$\ or $\Xi$\ system, and it does not affect the dip depth of reflection spectra. Additionally, EIT is contributed from the atoms with positive velocity when probe and pump fields are in a counter-propagating configuration.

In conclusion, the pump Rabi frequency and coherence decay rate play the main role in EIT at the interface. The probe Rabi frequency and the initial state after the collision between atoms and the interface have little effect on EIT. This study may pave the way for further investigation of quantum coherence and dynamics processes in the vicinity of the interface \cite{Thomas}.

\section*{ACKNOWLEDGMENT}

The authors thank Dr. D. Sheng for helpful discussions. The work was supported by the 973 Program (Nos. 2012CB921603), Natural Science Foundation of China (Nos. 61275209, 11304189, 61378015, and 11434007), NSFC Project for Excellent Research Team (No. 61121064), and PCSIRT (No. IRT13076).

\end{document}